\documentclass[11pt]{article}
\usepackage[margin=3cm]{geometry}
\usepackage{amsmath, amsthm}
\usepackage{natbib}
\usepackage{colortbl}
\usepackage{amsmath}
\usepackage{epsfig}
\usepackage[all]{xy}
\newcommand{\ci}{\mbox{\protect $\: \perp \hspace{-2.3ex}\perp$ }}
\newcommand{\tr}{^{\prime}}
\def\b#1{\mbox{\boldmath $#1$}}    

\newcommand{\diag}{{\rm diag}}    
\newcommand{\Ex}{\mbox{E }} 
\newcommand{\V}{\mbox{Var }}

\title{Modelling sources of
ecological fallacy within a revised Brown and Payne model of voting
transitions}
\author{A. ~Forcina,\\
Dipartimento di Economia, Finanza e Statistica, University of
Perugia, Italy,\\email: {\it forcina@stat.unipg.it}}
\begin{document}
\maketitle
\begin{abstract}
We present a model of voting behaviour based on a version of
aggregated overdispersed multinomial distributions; relative to a
similar model by \citet{BP86}, our model is based on more realistic
assumptions and free from certain shortcomings of the previous
model. We show that, within this model, it is possible to test for
certain confounding effects due to observable covariates measured at
the aggregate level; such effects, if ignored, might cause
substantial bias in the estimated relation between voting decisions
in two close in time elections, a phenomenon known as {\em
Ecological Fallacy}. An application to a referendum following an
election for the major in the town of Milan, which was interpreted
as a defeat for the Berlusconi gouvernment, is used as an
illustration.
\end{abstract}
\section{Introduction}
In its simplest form, ecological inference aims to draw conclusions
about the joint distribution of a pair of variables measured on
subjects belonging to a given population, from data aggregated
within geographic units, see for instance \citet{GreQui}. These
procedures are also known as ecological regression in the simpler
context where a linear regression model is assumed to holds at the
individual level as in \cite{GelPar}. The fact that the strength and
even the direction of the dependence at the aggregated level may be
substantially different from that at the individual level, a
phenomenon known as {\em Ecological Fallacy}, is well known since
\citet{Robinson}. The conceptual advantages of deriving a model for
the aggregated data from a well specified model at the individual
level has been emphasized by \citet{GutShe}, see also the additional
references they provide. They describe in details a collection of
fictitious examples that illustrate the problems that can arise by
neglecting the effects of aggregation. The bayesian model of
\citet{Wake} for the case of data in the form of a $2\times 2$ table
and that by \citet{GreQui} for the general $r\times c$ case may be
seen as well specified models at the level of individuals which,
among other things, help to clarify the sources of biasses that may
be introduced by aggregation.

\citet{Ki97} and \citet{GlyWak}, among others, have discussed
application of ecological inference to assess the effect of race on
voting behaviour. The model by \citet{BP86}, which we will discuss
in more detail below, is also a model of voting behaviour specified
at the level of individuals. Estimates of voting transitions between
two successive elections are relevant to political scientists who
may want to know if, for instance, the victory of a candidate was
due to her/his ability to attract voters who abstained in the
previous election or to {\em strategic vote} by supporters of a
minor candidate who wanted to prevent the main opponent from wining
\citep[see for instance][]{Alvar}. Estimation of voting transitions
is a natural field of application of ecological inference because
the data on the number of votes cast on each option recorded at the
level of polling stations are freely available in most European
countries. On the other hand, survey data where voters are asked
what they did in the two last elections, in addition to being
expensive to collect, are biased because voters may not tell the
truth or refuse to respond; in addition, non voters would be
difficult to include .

The approach to likelihood inference on voting transitions presented
in this paper build on the work of \cite{FoMa} and is an attempt to
improve the model of \citet{BP86} by assuming a different structure
of over-dispersion which we believe provides a more realistic model
of voting behaviour and, in addition, removes a technical problem
related to the central limit approximation of the likelihood. In
section 2 we describe the nature of electoral data, recall
\citet{BP86} model, formulate our proposal and discuss, briefly,
likelihood based versus Bayesian inference in this context. In
section 3 we examine certain possible sources of ecological biases
in the context of voting transitions that can be embedded, and thus
estimated, within our model; we also report the results of a small
simulation study documenting the performance of the estimation
method in such contexts. In section 4 we discuss an application to a
recent election in the city of Milan, Italy.
\section{Models of voting behaviour}
To introduce the subject, it may be useful to review briefly the
formulation of \citet{Good}. Suppose that the region of interest is
divided into $k$ polling stations; let $X$, $Y$ denote,
respectively, the choices made in two successive elections and
$$
P(j;i,s) = pr(Y=j\mid X=i, S=s),\quad j=1,\dots c, i=1,\dots, r,
s=1,\dots k
$$
the probability that a voter in polling station $s$ chooses option
$j$ in the latest election, having chosen $i$ in the previous one.
Suppose we assume that
\begin{enumerate}
\item
transition probabilities do not depend on the polling station, that
is $P(j;i,s)=P(j;i)$;
\item
voters decide independently of one
another.
\end{enumerate}
Let $\b n_s$ = $(n_{1s}, \dots , n_{rs})\tr$ denote the vector
containing the number of votes for each option at the earlier
election in polling station $s$ and $\b y_{is}$ = $(y_{1is}, \dots ,
y_{cis})\tr$ the number of votes for the options available at the
latest election among voters in polling station $s$ who had chosen
option $i$ in the earlier election. If $\b y_{is}$ was observable,
the above assumptions imply that
$$
\b y_{is} \sim Multinomial(n_{is}, \b p_{i}),
$$
where $\b p_{i}$ = $(P(1;i), \dots , P(c;i))\tr$. These data could
be arranged in a two way contingency table within each polling
station with the voting options  of the earlier and latest election
arranged by row and column respectively; however, only the row and
column marginals $\b n_s$ and $\b y_s$ = $\sum_i \b y_{is}$ are
available. By taking expectations within each multinomial, it
follows that $\Ex(\b y_s\mid \b p_1,\dots ,\b p_r)$ = $\sum_i
n_{is}\b p_{i}$, hence the transition probability may be estimated
by linear regression. To do so, we need to remove the last entry
from each $\b y_s$ to account for the fixed row totals and stack the
resulting vector of frequencies one below the other. This provides
$k(c-1)$, observations to estimate the $r(c-1)$ parameters; thus the
model is identified when $r \leq k$.

Though linear regression provides unbiased estimates of transition
probabilities under the assumption of multinomial distribution,
these are inefficient because the covariance structure is ignored;
in addition, estimated transition probabilities  may not lay between
0 and 1, in which case adjustments are required.
\subsection{The aggregated compound multinomial model}
The contribution of \citet{BP86} may be seen as a substantial
improvement both in the formulation and in the method of estimation
relative to the previous model. They start by assuming that the
behaviour of individuals, like in the Bayesian models to be
discussed below, is determined by vectors of transition
probabilities $\b p_{is}$ which are specific to each polling
station; however, to make the model identified, they assume that
variability across polling stations which is not accounted by
covariates is random, more precisely
$$
\b p_{is} \sim Dirichlet(\b \pi_i,\theta_i),
$$
where $\Ex(\b p_{is})$ = $\b\pi_i$ and $\V(\b p_{is})$ =
$\theta_i[\diag(\b\pi_i)-\b\pi_i\b\pi_i\tr]$. This random effect
model is equivalent to assume that the behaviour of the $n_{is}$
voters living within the boundaries of the same polling station, who
made the same choice in the previous election, are correlated,
probably because they share similar local features which are,
usually, unobserved.

Brown and Payne also worked out the full likelihood; parameterized
with the logits of the expected transition probabilities and those
of over-dispersion
$$
\lambda_{ij}=\log\frac{\pi_{ij}}{\pi_{ic}},
\:\tau_i=\log\frac{\theta_i}{1-\theta_i},  \:i=1,\dots, r, j=1,\dots
,c-1.
$$
Because the likelihood involves the sum of a product of multinomials
and is very hard to compute, unless the row totals in $\b n_s$ are
very small, they suggested a central limit approximation. Maximum
likelihood estimates are computed by maximizing the approximate
likelihood with respect to the parameters which, in the simplest
case of no covariates, are $\lambda_{ij}$ and $\tau_i$ parameters
and the transition probabilities are obtained by the reconstruction
formulas
$$
\hat{\pi}_{ij}=\frac{\exp(\hat\lambda_{ij})}{1+\sum_1^{c-1}
\exp(\hat\lambda_{ij})}, \: \hat\pi_{ic}=1-\sum_1^{c-1}
\hat{\pi}_{ij}.
$$
The logit link used in the model insures that the estimated
transition probabilities are strictly positive and sum to 1. In
addition, it provides a natural framework for modelling the effect
of covaruates measured at the level of polling stations; this is
discussed in detail below. \citet{FoMa89} proposed some minor
extension of the model and gave more details on computation of the
score vector and information matrix.
\subsection{The new model}
One problem with the above model is that the compound multinomial,
that is the distribution obtained by integrating out the $\b
p_{is}$, has a variance function of the form
$$
\V(\b y_{is}\mid \theta_i,\b\pi_i)=n_{is}[1+\theta_i (n_{is}-1)]
[\diag(\b\pi_i)-\b\pi_i\b\pi_i\tr]
$$
which is quadratic in the sample size. Intuitively, under the
assumed model each $\b y_{is}$ is affected by a single draw from a
Dirichelet, irrespective of the sample size; because of this the
central limit approximation is likely to be inaccurate. On a more
substantive ground, it seems more realistic to assume that voters
interact with each other within smaller circles whose size can vary
at random and should not increase with the sample size. With this in
mind, \citet{FoMa} propose the following assumptions for modelling
over-dispersion:
\begin{itemize}
\item
voting behaviour in the latest election of voters of party $i$ in
the earlier election who live in local unit $s$ is homogenous within
small clusters and is determined by a vector of transition
probabilities which is specific to cluster $h$ within polling
station $s$, these probabilities are denoted $\b p_{ish}$ and we
assume that
$$
\b p_{ish}\sim Dirichlet(\b\pi_i,\theta_i);
$$
\item
the sizes of the clusters, within which voters of party $i$ in
polling station $s$ split, follows a Multinomial$(n_{is}, \b 1
(C_{is}/n_{is}))$, where $C_{is}$ is the average cluster size; note
that this implies that clusters are, on average, of the same size.
\end{itemize}
In the Appendix we show that, under these assumptions
\begin{equation}
\V(\b y_{is}\mid \theta_i,\b\pi_i, C_{is}) = n_{is}\left[1+\theta_i
C_{is}\frac{n_{is}-1}{n_{is}}\right]
[\diag(\b\pi_{i})-\b\pi_{i}\b\pi_{i}\tr]. \label{eq:var}
\end{equation}
This expression is linear in the sample sizes $n_{is}$ for fixed
$C_{is}$, the average cluster size which should be reasonably small.
Because $C_{is}$ and $\theta_i$ are not identifiable simultaneously,
we propose to assume that $C_{is}$ = $C$ is constant and known and
then to perform a sensitivity analysis to assess the effect of
different choices for $C$. Our sensitivity analysis indicates that
changing the value of $C$ has no detectable effect on the estimates
of transition probabilities and only a minor effect on the estimate
of $\theta_i$.

Under the normal approximation, the likelihood depends on the
parameters that determine the transition probabilities $\pi_{ij}$ on
the logit scale and on the over-dispersion parameters $\theta_i$
which is also parameterized on the logit scale. The log-likelihood
can be maximized by a Fisher scoring algorithm, this requires
computing the score vector and the expected information matrix. At
convergence, the information matrix may be used to compute
asymptotic standard errors by a first order approximation. Because
these calculations are rather tedious and the tools required are
essentially the same outlined by \citet{BP86}, they will be made
available as supplementary material.
\subsection{Bayesian versus Likelihood inference}
The point of view taken in this paper is that Bayesian or likelihood
based inferences should be assessed on their own grounds and may be
both valid. The likelihood assumed in the Bayesian models proposed
by \cite{Wake} for the $2\times 2$ case and by \cite{GreQui} for the
general $r\times c$ case is identical to the ones assumed here and
in the Brown and Payne model. However, the quantities of interest
are usually different: in the Bayesian approach one obtains the
posterior distribution of the cell frequencies in each local units
conditionally on the observed margins. Instead, likelihood inference
aims at estimating the transition probabilities $\b\pi_i$, $i=1,
\dots r$ for the whole area, in additiona to regression coefficients
if covariates are available. An obvious property of the Bayesian
estimates of cell frequencies is that they are always consistent
with the rows and columns totals, a property whose importance has,
perhaps, been overemphasized in the work of G. King, see for
instance \cite{Ki97}. Estimates of transition probabilities at the
level of local units or for the whole population may then be easily
computed if of interest.

An advantage of the likelihood approach is that its results are not
dependent on the choice of an appropriate hierarchical prior.
\cite{GreQui}, among others, have investigated the sensitivity of
the results to the prior specifications.  From a non-Bayesian point
of view, the estimated transition probabilities need not be exactly
consistent with the marginal frequencies: the objective of inference
is to estimate the parameters that determine the assumed generating
mechanism, not the actual number of transitions. On the other hand,
the problem of estimating cell frequencies conditionally to the
observed marginals and the estimated parameters, can be framed as
the problem of computing the expected value of a multivariate
extended hypergeometric distribution, see \cite{BFD:01}, section
3.3. An early attempt to solve this problem is due to \cite{Firth}
who was trying to implement an EM algorithm for ecological
inference. Though there is no closed form expression for computing
this expectation, good approximations may be provided by the
iterative proportional fitting algorithm or by a normal
approximation of the extended hypergeometric.
\section{Ecological bias and the effect of covariates}
The nature of the ecological fallacy and its relation with the
Simpson's paradox is clearly explained by \citet{Wake}, section 3.3,
in the context of dichotomous variables. A general discussion may be
based on directed acyclic graphs (DAGs) with a causal
interpretation; within this framework, the direct effect of $X$ on
$Y$ is determined by the transition probabilities. In Fig. 1 below,
the dependence of $Y$ on $X$ can be recovered without bias from
aggregated data only in the causal model on the left, where $Y \ci
S\mid X$ like in the basic Goodman model. The result follows from
the fact that the conditional probabilities $P(Y=j\mid X=i)$ are the
unknowns in the system of linear equations that can be be derived
from the DAG
$$
P(Y=j\mid S=s) = \sum_i P(Y=j\mid X=i) P(X=i\mid S=s);
$$
this system has a unique solution if the equations are consistent
and, like for the Goodman model, $r\leq k$. The other two graphs to
the right characterize contexts where ecological fallacy is
possible, essentially because $X$ and $Y$ have a common "cause"
inducing confounding. When this happens, data on the joint
distribution of $X,\:Y$ and the confounder is necessary to obtain an
unbiased estimate of the effect of $X$ on $Y$; this follows, for
instance, by applying the {\it do} operator in \cite{Pear95}.
\begin{figure}[h]
\begin{center}
$$
\xymatrix{
 S \ar@{->}[d] &  & & S \ar@{->}[d]
\ar@{->}[dr] &  & & S \ar@{->}[d] \ar@{->}[r] & V \ar@{->}[d] \\
 X \ar@{->}[r] & Y & & X \ar@{->}[r] & Y & & X \ar@{->}[r] & Y
 }
$$
\caption{Three types of causal DAGs, only the first on the left
cannot lead to ecological fallacy.}
\end{center}
\end{figure}
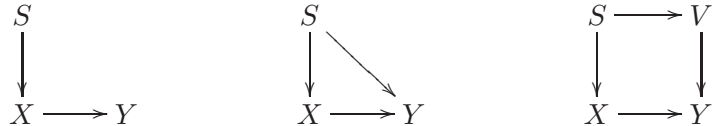
The second DAG of Figure 1 assumes that the nature of the local
units affects voting decisions in both elections; this may happen,
for instance, when there are areas with a strong tradition of right
(or left) wing affiliation: it is likely that those who voted a
given party in the previous election may behave differently
depending on whether or not they live in areas where that party has
a strong local roots. This context is similar the one described by
\citet{GutShe} where $X$ denotes the event of being protestant and
$Y$ that of committing suicide: a simple ecological regression
indicated positive association leading to the conclusion that
suicide was more common among protestants. A more detailed analysis
revealed that living in an area with a protestant majority caused
distress among catholic residents leading to a greater number of
suicides among catholics. In the last DAG on the right, local
features affect a variable $V$ which has an influence on the results
in the last elections. For instance, $V$ might be the party voted at
an even earlier election or the age group of the voter.

A special case of the second DAG above is when the proportion of
voters with a given value of $X$ may be used as a covariate to
control for confounding. \citet{ChaSte} explain why, within a linear
regression model like Goodman's, this additional effect would not be
identifiable.  Our model, as well as Brown and Payne's logistic
model, is free from this problem because the dependence of the
transition probabilities on $X$ is not linear on the logistic scale.
The model depicted in the third DAG is more problematic because to
model the distribution of $Y\mid X,V$, observations on the marginal
distributions of $(S,X,V)$ and $S,Y$ would be required. A model
where we use the proportion of voters for a given options of $V$ as
a covariate might help removing bias only in very special
circumstances. Suppose, for instance, that $V$ is the choice made in
a previous election of a different kind; it might be that the local
units with a larger proportion of voters for party $V=v$ in that
election are those where the party has a larger number of militants
and stronger roots, a feature which might affect the behaviour of
local voters. However, this model is structurally different from a
model where we assume that, for instance, those who voted party $v$
in two previous elections are more likely to vote it again relative
to those who voted it only in the previous election. To estimate
such a model by ecological inference, data on the joint distribution
of $V,\:X$ in each polling station wold be required.

Whenever confounding may be controlled, at least approximately,
bythe introduction of appropriate covariates as described above,
these may be inserted in the logistic link function; for instance,
for a single covariate $V$, let
\begin{equation}
\log(\pi_{sij}/\pi_{sic})=\alpha_{ij}+\beta_{ij}v_s,
 \label{lcov}
\end{equation}
where $\alpha_{ij}$ denotes the baseline logit and $v_s$ is a known
function of the frequency distribution of $V$ in polling station
$s$. Note that, when inserting a single covariate into a model of
vote transitions, unless some restriction on the $\beta_{ij}$ is
imposed, $r(c-1)$ additional parameters are required. More
parsimonious models involving covariates may be constructed by
assuming, as we do in the application, that a given covariate
affects only a given cell so that all the $\beta_{ij}$ are 0 except
one.
\subsection{A simulation study}
The model described in Section 2 may be seen as a data generating
mechanism that can be implemented to produce electoral data from a
given probability distribution. We first use this device to show how
a specified amount of confounding may distort the linearity of
regression predicted in the Goodman model and then to assess the
performance of our estimation algorithm when the effect of
confounding can be modelled by the inclusion of covariates.
Generation of artificial electoral data are based on the following
steps:
\begin{itemize}
\item
use a uniform within preassigned bounds to sample the number of
voters and the probabilities of voting for the available parties in
the first election in each polling station and then a multinomial to
obtain the actual number of votes in the latest election;
\item
for $s=1,\dots k$ and $i=1, \dots, r$, the number of clusters is set
equal to the integer part of $n_{is}/C$, where $C$ is the average
cluster size, the sizes of these clusters are sampled from a
multinomial with uniform probabilities;
\item
the actual transition probabilities within each cluster are sampled
from a Dirichelet whose expected values may depend on covariates;
these probabilities are used to sample the actual number of votes
for the options  available in the second election within each
cluster; the final results in the second election are obtained by
summing across clusters and parties.
\end{itemize}

For simplicity, we assumed that only two options were available in
both elections. In (\ref{lcov}) let $\alpha_{11}=\log(.7/.3)$,
$\alpha_{21}=\log(.2/.8)$ and  $v_s$ = $x_s-\sum_s v_s/k$, where
$x_s$ = $\log(n_{s1}/n_{s2})$, that is, the only available covariate
is the centered logits of the results in the first election.
Consider three alternative models: (i) no confounding, that is
$\beta_{11}=\beta_{21}=0$, (ii) concordant effects with
$\beta_{11}=1$, $\beta_{21}=0.5$, (iii) discordant effects with
$\beta_{11}=-1$, $\beta_{21}=0.5$. Data for 200 polling stations
having between 600 and 800 voters were generated with
$\theta_1=\theta_2=0.1$; the proportion of votes for party 1 in both
elections are plotted in Fig. 2. The results of an ecological
regression which neglected confounding would fit a regression line
to the observed points: in (i) the observed points follow the true
regression line; under (ii) the observed regression line is much
steeper, leading to over estimate the association between $X$ and
$Y$; under (iii) the observed relation is non linear and the overall
association appears slightly negative, leading to a substantial
underestimation of the true association.
\begin{figure}[ht]
\begin{center}
 \includegraphics[width=16cm,height=6cm]{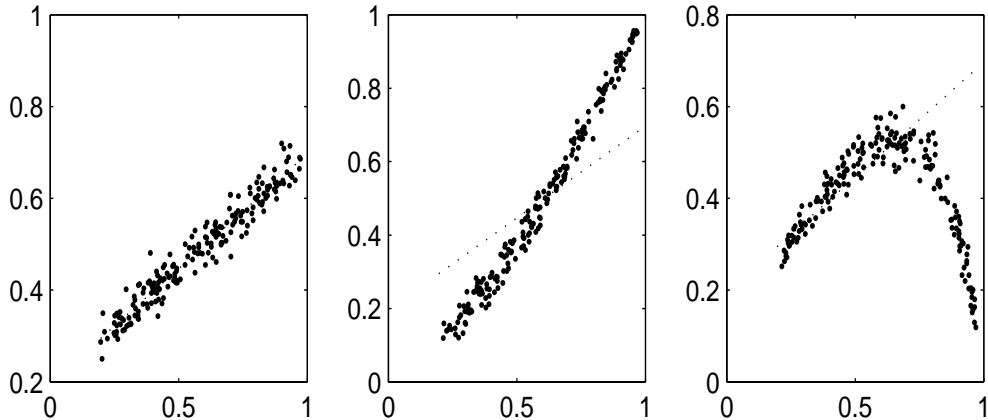}
 \caption{Simulated proportions of voters for party 1 in the first
 (X) and the second (Y) election for no confounding, concordant and
 discordant effects; dashed line is the expected relation under no confounding.}
\end{center}
\end{figure}

Under the discordant model and the settings described above, 1000
samples were generated and, for each sample, the maximum likelihood
estimates under the correct model were computed. To see how the
properties of the estimates changed with the sample size, estimates
were computed also for the case where the number of voters per
polling stations were between 2400 and 3200. Estimates of the bias
on each parameter are given in Table 1 below: they are small and
seem to decrease with sample size. Only for the overdispersion
parameter, the bias is a little more substantial. Ratios between the
average standard errors of the estimates computed from the
information matrix and those estimated from the sample variances are
given in the same table: the fact that the ratios are smaller than 1
indicates that standard errors computed from the information matrix
underestimate slightly sample variability and the effect decreases
with the sample size. Again the performance is worst relative to
$\tau$.
\begin{table}[ht]
\caption{\it Sample bias and ratios of standard errors based on
information matrix and sample variances } \label{tab:1}
\begin{center}
\begin{tabular}{llrrrrc} \hline
Sample size &  & $\alpha_{11}$ & $\alpha_{21}$ &
$\tau$ & $\beta_{11}$ & $\beta_{21}$\\
 \hline
600-800 & Bias & -0.0202 & 0.0312 & -0.1006 & 0.0054 & 0.0118\\
2400-3200 & Bias & -0.0038 & 0.0069 & -0.0744 & 0.0011 &  0.0028\\
600-800 & Ratios of s.e. & 0.9654 &  0.9616 &  0.9140 &  0.9656 &  0.9641\\
2400-3200 & Ratios of s.e. & 0.9943 & 0.9985 & 0.9330 & 0.9939 & 0.9906
\\  \hline
\end{tabular}
\end{center}
\end{table}
The proportion of sample estimates with an absolute error that
exceeds given thresholds taken from the normal distribution are
displayed in Table 2: when the sample size is smaller the
distribution of the sample estimates for most parameters appear to
have heavier tails relative to the normal.
\begin{table}[hb]
\caption{\it Proportion of sample estimates whose absolute error
exceeds $z\hat{\sigma}$ } \label{tab:1}
\begin{center}
\begin{tabular}{lrrrrcrrrr} \hline
$z$ & 1.2815 & 1.6449 & 1.9600 & 2.576 & \hspace{1mm} & 1.2815 &
1.6449 & 1.9600 & 2.576 \\ \hline
 $\alpha_{11}$ & 0.1660 & 0.0960 & 0.0500 & 0.0200 & & 0.1960 &
 0.0920 & 0.0470 & 0.0140\\
 $\alpha_{21}$ & 0.1780 & 0.1040 & 0.0510 & 0.0150 & & 0.2010 &
 0.0910 & 0.0490 &  0.0120\\
 $\tau$ & 0.1890 & 0.1090 & 0.0540 & 0.0140 & & 0.1860 &
 0.1020 & 0.0570 &  0.0160\\
 $\beta_{11}$ & 0.1800 & 0.0850 & 0.0540 & 0.0190 & & 0.1880 &
 0.0920 & 0.0450 &  0.0100\\
 $\alpha_{21}$ & 0.1950 & 0.1070 & 0.0560 & 0.0140 & & 0.1950 &
 0.0980 & 0.0490 & 0.0160 \\
 \hline
\end{tabular}
\end{center}
\end{table}
\section{An application}
In May 15th 2011 in the city of Milan there was an election for the
mayor and party representatives in the city council followed on the
29th by a run-off ballot between the previous mayor Mrs Moratti
supported by the Mr Berlusconi's gouverning coalition and Mr Pisapia
supported by most parties from the left. A national referendum
promoted by the opposition party Italia dei Valori (IdV) was held on
June 12th to cancel several laws passed by the gouvernment. One of
these laws allowed the Prime Minister, who is under trial under
several accusations, to claim that the duties of his office
prevented him from attending the Courts, thus postponing a decision
at his will. In this application we study the behaviour of voters at
the Referendum on this particular issue relative to their behaviour
at the run-off ballot: Berlusconi and his supporters had campaigned
for not going to vote at the referendum because, if less than 50\%
went to vote, the result of the referendum would be irrelevant. Lega
Nord (LN), which belongs to the gouverning coalition, was much less
motivated in supporting Berlusconi on this specific issue relative
to voters of Partito delle Liberta' (PdL), Berlusconi's own party.
It is likely that some of the supporters of the center opposition
party Unione di Centro (UdC), who abstained at the run-off feeling
that Pisapia was much to the left, decided to vote with the left
parties at the referendum.

The municipality of Milan has about 990 thousands voters; after
excluding the few atypical polling stations like hospitals, we could
use data from 1159 polling stations. As covariates we used the
centered logits of the proportion of votes cast on May 15the for the
following options: (i) Pdl, (ii) (LN), (iii) UdC, (iv) (IdV), (v)
abstain and blank ballots (NoV). The estimated transition matrix
with standard errors, averaged across polling stations, are given on
the left-hand side of Table 3 and the estimated effects of
covariates are on the right-hand side of the same table. As
expected, a large proportion of those who supported Pisapia voted
"yes" at the Referendum which was also supported by almost 30\% of
voters who had abstained at the run-off. Even more interesting is
that over 22\% of those who had supported Moratti did not follow
Berlusconi's recommendations and went to vote at the Referendum,
Estimates of the regression parameters, which are all highly
significant is briefly discussed below. Note that, whenever we
assume that a covariate affects the residual category of NoV, the
actual sign of the regression parameter which is estimated from the
induced effect on the other voting options, is reversed.
\begin{table}[ht]
\caption{\it Average estimated transitions from Ballot to Referendum
and covariates effects } \label{tab:1}
\begin{center}
\begin{tabular}{lrrrclrr} \hline
\multicolumn{4}{c}{Transition Probabilities} & \hspace{1mm} &
\multicolumn{3}{c}{Regression estimates} \\
 & Yes &  No & NoV & & Covariate & $\hat{\beta}$ & s.e.($\beta$) \\
 \hline
Moratti & 0.1388 & 0.0858 & 0.7754 & & LN on Yes $\mid$ Moratti &     0,3311 & 0,1061 \\
s.s. & 0.0208 & 0.0068 & 0.0219 & & UdC on Yes$\mid$ Pisapia & 0.0810 & 0,0282 \\
Pisapia & 0.9380 & 0.0000 & 0.0620 & & NoV on Yes-No $\mid$ NoV & -0.7326 & 0,0692 \\
s.e. & 0.0140 & 0.0000 & 0.0140 & & PdL on Yes-No $\mid$ Moratti & -0.3996 &   0,0637 \\
NoV & 0.2938 & 0.0428 & 0.6633 & & IdV on Yes-No $\mid$ Pisapia & 0.4167 & 0.078 \\
s.e. & 0.0211 & 0.0061 & 0.0223 & & & & \\
 \hline
\end{tabular}
\end{center}
\end{table}
\begin{itemize}
\item
the proportion of those who voted "yes" at the Referendum, having
supported Moratti at the Ballot, increases in polling stations with
a larger proportion of voters for LN; this indicates that these
voters were reluctant to follow Berlusconi on this issue;
\item
the proportion of those who abstained at the Referendum, having
supported Moratti at the Ballot, increased in polling stations with
a larger proportion of voters for PdL; this indicates that PdL
supporters tended to obey the recommendation of their party;
\item
the proportion of those who voted "yes" at the Referendum, having
abstained at the Ballot, increased in polling stations with a larger
proportion of voters for UdC as expected;
\item
the proportion of those who abstained at the Referendum, having
abstained at the Ballot, increased in polling stations with a larger
proportion of abstainers; this indicates that there is a structural
proportion of voters who abstain most of the time;
\item
the proportion of those who abstained at the Referendum, having
supported Pisapia at the Ballot, decreased in polling stations with
a larger proportion of voters for IdV; this is consistent with the
fact that these voters were the most militant on the issue.
\end{itemize}

\end{document}